\documentclass[twocolumn,aps,prl,epsfig]{revtex4}

\usepackage{epsfig}
\usepackage{dcolumn}
\usepackage{bm}
 
\begin{document}

\title{Remarks on the first hundred years of superconductivity}

\author{Karol Izydor Wysoki\'nski} 

\affiliation{Institute of Physics,
M. Curie-Sk\l{}odowska University,
Radziszewskiego 10, 
Pl 20-031 Lublin}


\date{ \today}

\begin{abstract}
On the occasion of centenary of superconductivity discovery 
I recall some facts from the first period and attempts to understand 
the phenomenon. It turns out that most famous physicists of the 
first half of XX century have tried to solve the puzzle.
 Bardeen, Cooper and Schrieffer succeeded in 1957. The BCS theory 
 successfully described all known facts and offered new predictions, 
 which soon have been confirmed experimentally contributing to the 
 widespread acceptance of the theory. 
 It have found applications in nuclear physics, theory of neutron stars
 and cold atomic gases.
 The discoveries of new superconductors in the last thirty 
 years  show that simple BCS model is not enough to understand 
 new unconventional superconductors.   
 The studies of superconductors  develop vividly and 
still  fascinate new generations of physicists working in such diverse
 fields as material science and string theory.
\end{abstract}
\maketitle
The subtitle of the present {\it XVth National School on High Temperature 
Superconductivity} is "Hundred years of superconductivity".
It is thus legitimate to recall some important facts from the first hundred years
of studies of the phenomenon of superconductivity.
The discovery of superconductivity in 1911 was unexpected, contrary to
two years earlier liquefaction of helium. Both these great 
achievements were reached in the Leiden Laboratory headed by
Heike Kamerlingh Onnes.
 
Heike Kamerlingh Onnes studied the properties of metals at low temperatures
not because anybody has expected or predicted the appearance 
of unusual phenomena there. The mere motivation was to answer simple
and controversial at the time questions of whether the resistance 
of metals will vanish as was expected by Drude theory. 
Paul Drude argued that at very low
temperatures the scattering of electrons by lattice vibrations will be
ineffective and the resistivity will decline with lowering temperature. 
On the other hand the speculative theory of Lord Kelvin (Wiliam Thomson), a very 
influential scientist, predicted that the resistivity should increase
(in fact to infinity) at low temperatures as a result of freezing of electrons 
to the atoms in the material. 

The vanishing of resistivity of mercury at low temperatures 
measured by Gilles Holst, one of the Kamerlingh Onnes' 
collaborators in Leiden was  thus for Kamerlingh Onnes 
 a natural and expected result. He was so sure of the result that 
  overlooked the sudden drop of the resistance
and finished the first report \cite{onnes1911} 
on low temperature study of resistivity in mercury saying 
that the behavior is {\it "in agreement with the theory"}. 
Only after further studies {\it inter alia} 
of mercury doped with gold  Kamerlingh Onnes realized 
that the completely new phenomenon is in play.

At this point it is worth to cite the remark 
Kamerlingh Onnes made on the same day,
8$^{th}$ of April 1911, in the laboratory notebook \cite{vandelft2010}: 
{\it "Just before the 
lowest temperature [about 1.8 K] was reached, the boiling suddenly stopped 
and was replaced by evaporation in which liquid visibly shrank. 
So, a remarkably strong evaporation at the surface"}. There is no doubt, that
the superfluidity of helium was also observed on that day. But nobody
in Leiden has paid any attention to that puzzling behavior of 
helium at low temperatures. The phenomenon of superfluidity in 
$^4$He was discovered in 1937 \cite{kapitza1937}, while the corresponding effect 
in lighter isotope $^3$He had to wait till 1972 \cite{osheroff1972}.

\begin{figure}[htb]
\includegraphics[width=20pc]{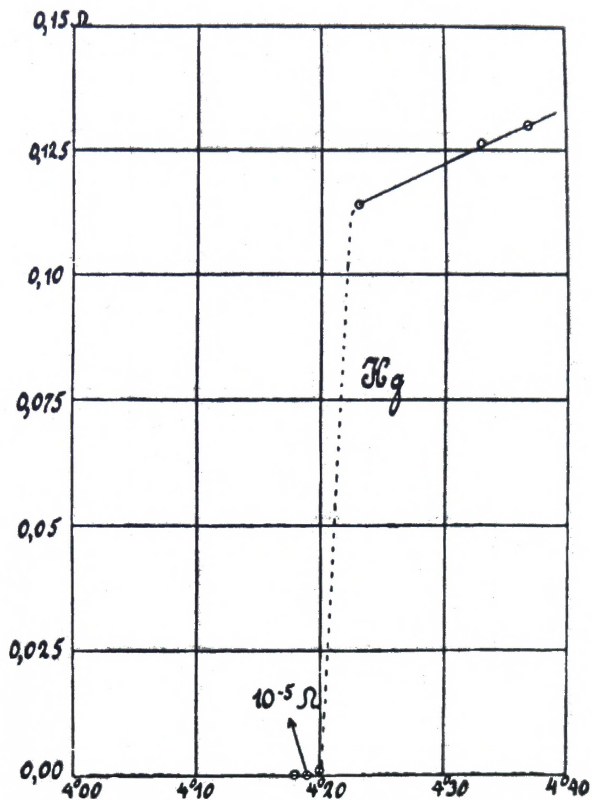}
\caption{\label{scHg} The  original plot 
showing the superconducting transition in Hg \cite{HOnnes1911b}.}
\end{figure}

In the Fig. (\ref{scHg}) the original data on temperature dependence of the
resitivity in mercury measured in Leiden  by Gilles Holst are reproduced. 
The figure 
was obtained   in October 1911 and published by
H. Kamerlingh Onnes without those who really conducted experiments.
At that time it was clear that a sudden drop of resistivity has nothing 
to do with the ineffectiveness of phonons and marks the appearance 
of the completely new phenomenon of "supraconductivity". 

H. Kamerlingh Onnes was a very practical man and 
starting the  studies of resistivity R in metals at helium temperatures 
he had in mind - besides checking if Kelvin's theory is correct - another
 idea  of using the linear temperature dependence
of R(T) to build new thermometer. Even before liquefaction
of helium he has observed linear behavior of R(T) of many metals
and basing on this linearity  wanted to construct an instrument alternative to 
gas pressure thermometers in common use at the time \cite{HOnnes1911b}.

H. Kamerlingh Onnes got the Nobel prize in 1913   \cite{onnes_nobel_l}
 {\it ``for his investigations on the properties 
of matter at low temperatures 
which led}, inter alia, {\it to the production of liquid helium".}
In the Nobel lecture \cite{onnes_nobel_l} he discussed mainly 
methods which anabled to reach the goal \cite{krak} and the properties
of gases but spent some time showing the resistivity of metals at low
temperatures and the appearance of superconductivity in mercury.

\begin{figure}[htb]
\includegraphics[width=20pc]{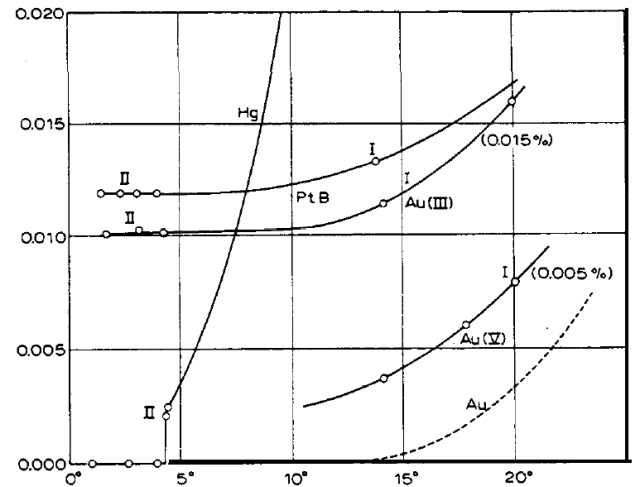}
\caption{\label{sc-metals} The  Kamerlingh Onnes  plot from his Nobel lecture
showing the low temperature measurements of the 
resistance in various metals (Au of different purity, Pt and Hg).
The dashed curve shows the expectations for the pure gold. The original
text describes the axis of the figure in the following way: 
{\it "The resistance, in fractions of the resistance at zero Centigrade, 
is shown as the ordinate and the temperature as the 
abscissa." \cite{onnes_nobel_l}}}
\end{figure}

In the Fig. (\ref{sc-metals}), taken from his Nobel lecture,  
the dependence of the resistance is shown as function of temperature
down to about 1.5 K, roughly the lowest temperature that could be obtained 
in Leiden Laboratory. Discussing the curve for mercury
Kamerlingh Onnes writes: {\it As has been said, the experiment left 
no doubt that, as far as accuracy of
measurement went, the resistance disappeared. At the same time, 
however, something unexpected occurred. The disappearance did 
not take place gradually but} abruptly. {\it From 1/500 the 
resistance at 4.20K drops to a millionth part. At the lowest 
temperature, 1.50K, it could be established
that the resistance had  become less than 
a thousand-millionth part of that at normal temperature.} 

During the Nobel lecture Heike Kamerlingh Onnes
discussed many different studies performed in Leiden at low temperatures 
in cooperation with various visitors to the laboratory (e.g. {\it "Mme. Curie 
[came] to examine the penetrating radiation of radium"}). 
It is also worth mentioning 
the paragraph \cite{onnes_nobel_l} from his Nobel lecture 
on the problems he encountered during 
the attempts to obtain very high magnetic fields with help of 
the superconducting coils. Shortly after the discovery
of the new effect he realised that superconductors should carry large 
electric currents.  Kamerlingh Onnes thought of using this to produce
huge magnetic fields.  

Being a very important and well known scientist Kamerlingh Onnes 
invited to Leiden other great physicist, namely Albert Einstein. 
Perhaps it is of interest to recall that  Albert Einstein got interested
in the subject and  
has published a paper on superconductivity on the occasion of
forty anniversary  of Kamerligh Onnes  professorship in Leiden.
In the paper, which is now accessible in English \cite{einstein1922}
Einstein developed the idea that  
{\it "supercurrents are carried through closed
molecular chains where electrons undergo continuous cyclic exchanges"}.
  He also proposed an 
experimental verification of his theory. Namely, one has to take
two different superconductors, connect them and look if the 
resistance  of such a system in series will vanish. If the answer 
is affirmative then the theory is wrong as molecular chains can not
extend between  two different materials. Kameringh Onnes performed
such experiment and falsified the Einstein's theory even before
the paper appeared in print. 

Probably  Kamerlingh Onnes had no chance to measure,
accurately enough the alternating current, which  obviously flew
between his biased superconductors. This current is known as and a.c.
Josephson effect \cite{josephson1962} and has been measured 
\cite{anderson1963} in 1963, after 
it was predicted theoretically by the young student in Cambridge
\cite{anderson1970}. The successful measurements of the
Josephson current was commented by one of the co-authors of the 1963
paper in such a way \cite{anderson1970} {\it "We were able to see the
effect because three conditions were
satisfied: First, we knew what to look
for; second, we understood what we
saw. Both of these were the result of
our contact with Josephson. The third
condition was that we were confident of
Rowell's skills in making good, clean,
reliable, tunnel junctions."}

The many attempts to understand the phenomenon by the 
greatest minds of the
first half of the twenties century are well known and have been 
discussed on many occasions \cite{sc-history}. That is why I shall not
dwell on that. Instead  I want to spent some time
by discussing two experiments, which greatly contributed to the proper
understanding of the phenomenon and to the formulation of the correct
phenomenological  and later microscopic theory of superconductivity.

The first of them is the discovery in 1933 by Meissner and Ochsenfeld
that the magnetic induction vanishes inside the superconductor
\cite{meissner1933}. In the "Short Note" section of the journal the
authors describe their main findings on roughly half a page and summarised 
their two  experiments  saying \cite{trans} that:

{\it 1.When the superconducting transition temperature was reached,
the magnetic field lines have changed in the neighborhood of
the superconductor in such a way as if the permeability  0
or susceptibility ${-1\over 4\pi}$ were to be expected in the superconductor.

2. In the inner part of the long lead tube the magnetic field 
- despite the changes of it observed outside the tube according to the effect
described  in point 1 - [...] remained nearly unchanged."}

These findings which showed that the metal -- superconductor transition 
is a reversible phenomenon allowed to apply the thermodynamic ideas and the
conclusion that the entropy of the normal state is larger than that
of a superconducting one and that the superconducting state is more 
ordered, with lower symmetry in analogy to paramagnet -- ferromagnet
transition. Certainly these findings have provoked Landau, Bloch and others
to propose theories of superconductivity analogous  to those of the
ferromagnetism. 

Heinz and Fritz Londons were the first to understand the real
importance of the Meissner and Ochsenfeld observation and soon 
propsed a phenomenological theory \cite{londons}  which was able 
to account for the observations. Fritz London  correctly  argued that the
superconductivity is a manifestation of quantum laws  on a macroscopic scale
of meters or kilometers - he coined the phrase "macroscopic quantum coherence". 
This line of thinking was crowned with
celebrated Ginzburg -- Landau thermodynamic description \cite{ginzburg1950}
of superconductors.  
 
\begin{figure}[htb]
\includegraphics[width=20pc]{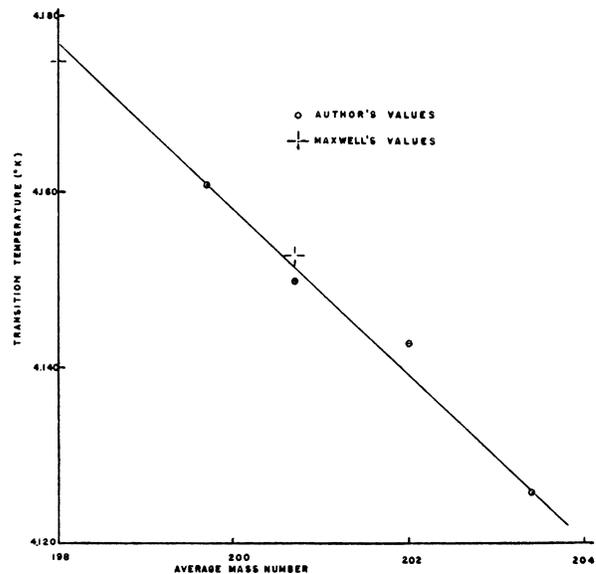}
\caption{\label{isotop} The  original plot \cite{reynolds1950} 
showing the isotop effect in Hg.}
\end{figure}

The other crucial experiment I want to mention is that reporting on the isotope
 effect. In fact there were two reports \cite{maxwell1950,reynolds1950}, 
 both obtained  by  Physical Review on the same 
 day 24 March 1950 and published in the same issue of
 the journal in section Letters to the Editor. One by E. Maxwell from NBS,
 Washington D.C. and other by C.A. Reynolds and Colleagues from Rutgers
 University, New Jersey. Both groups were aware of their work and mutually cited
 each others work. It is important to note the nice agreement of those very precise
 and carefull measurements as can be seen in the  Fig. (\ref{isotop}) reproduced
 from \cite{reynolds1950}. The superconducting transition temperature 
 of natural mercury with average atomic weight 200.6 was found to be 4.156 K
 in one the papers \cite{maxwell1950} and T$_c$=4.177 K for $^{198}$Hg. The other
 group \cite{reynolds1950} reported  T$_c$=4.150 K for natural mercury and 
 T$_c$=4.143 K for the sample consisting of the $^{202}$Hg isotope.
 The average slope of the line in the Fig. (\ref{isotop}) has been estimated 
 to be 0.009 K/(mass number).

 It is interesting to note, that only two month later the theoretical paper
 by H. Fr\"ohlich proposing that the
 interaction between electrons and lattice vibrations was responsible 
 for the superconductivity \cite{frohlich1950} was received by the same journal. 
 The author mentioned the 
 experimental papers on the isotope effect only in the "Note added in proof" 
 with the remark: {\it " The isotope effect [...] which has recently come to my
 notice follows immediately from the proportionality of $|S|$
 with the inverse isotopic mass 1/M [...]. This agreement provides a direct
 check for the fundamental assumptions of the theory."} 
 In his theory Fr\"ohlich finds an effective 
 {\it "interaction in the momentum space"}.
 As we know it from the BCS theory \cite{BCS} this effective attractive 
 interaction is responsible for the superconducting instability. In the
 paper \cite{frohlich1950} one reads about the efective interaction:  
 {\it "roughly speaking 
 it is repulsive for equal energies but different directions of $\vec k$ 
 and attractive otherwise"}. Both, the experimental data on the isotope effect
 and  the theoretical analysis of effective interaction, together with the
 prediction of its attractive character contributed greatly to the formulation of
 the correct microscopic theory of superconductivity \cite{BCS} in 1957.
 In this context it is important to mention 
 that Kamerligh Onnes and Tuyn in 1922
and indpendently E. Justi in 1941 (see citations 1 and 2 in
\cite{reynolds1950}) have checked that the superconducting 
transition temperature
of the natural lead is the same as that of lead obtained 
from the decay of uranium.

In the decade after the correct microscopic description of superconductivity
has been achieved \cite{BCS}  many new experiments were conducted which
fully supported the theory. The detrimental role of magnetic impurities 
on the superconductivity and their robustness against non-magnetic impurities
have been explained, respectively  by A.A. Abrikosov  and P.W. Anderson. 
The strong coupling approach to the superconductivity was developed by G. M.
Eliashberg and applied to  strong-coupled superconductors by W.L. McMillan. 
As a result in the late sixties of the last century 
the superconductivity research was in decline.
The famous two volume work \cite{parks1969} published in 1969 has summarised 
all known facts and theories of classic superconductors and was considered 
by some as a last nail to the coffin of superconductivity. 

The completly unexpected discovery of superconductivity with relativly high
transition temperature of about 9K in PdH$_x$ \cite{skoskiewicz1972} by prof.
Tadeusz Sko\'skiewicz in 1972  
and shortly after that the negative isotope effect \cite{skoskiewicz1974} 
by the same author gave 
new impetus to the field. The other great Polish discovery 
of the time, namely the
coexistence of superconductivity and ferromagnetism 
in $Y_9Co_7$ (or $Y_4Co_3$) 
by prof. Andrzej Ko\l{}odziejczyk \cite{kolodziejczyk1980} 
remained virtually unnoticed. 
The superconductivity in heavy fermion system Ce Cu$_2$Si$_2$ with T$_c$=0.5 K
was discovered in 1978 by Frank Steglich {\it et al.} \cite{steglich1979} 
and still provides a puzzle \cite{spalek1997}. 

The situation changed completely after the 
discovery of high temperature superconductivity (HTS) in La oxides by Bednorz
and M\"uller in 1986 \cite{muller1986}, followed by the discovery of 
other families of  HTS with maximum transition temperatures up to 138 K
at ambient  and 166 K at high pressure in HgBa$_2$Ca$_2$Cu$_3$O$_{8-\delta}$. 
These findings evoked an  unprecedented surge of experimental and theoretical 
activities in the field. We note by passing the 25$^{th}$
anniversary of HTS and the fifth anniversary of the first report
\cite{kamihara2006} on the 
superconductivity in iron pnictides - the other class of high temperature 
superconductivity with highest T$_c$=56K.

The numerous experimental discoveries of new superconductors 
during last three decades (strontium
ruthenates, alkali doped fullerides, MgB$_2$, 
non-centrosymmetric superconductors, etc.) have
sparked new and unprecedented research activity with many new 
theoretical ideas and views. Unconventional superconductors are 
at the forefront of condensed matter physics. The microscopic mechanisms
operating in those materials are at the center of scientific 
debate \cite{norman2011}.

The BCS pairing theory explaining the resistance-less flow of electrons accompanied 
by the expulsion of the magnetic field from inside of the superconductor
has influenced the way we think on various problems in physics. It has found many
application in different scientific areas {\it e.g.}: 
it was applied  (i) to atomic nuclei 
to explain why the even-even nuclei are more tightly bound than even-odd or odd-odd ones, 
(ii) to neutron stars which superfluidity may explain some properties of pulsars,
(iii) the pairing may be in play in quark-gluon plasma expected to appear 
during heavy ion collisions. 

The description of superfluidity in $^3$He at about 2 mK
 and fermionic ultra cold atoms at temperatures as low as 10$^{-8}$ K require pairing of
atoms. In the case of  $^3$He the quantum state  is an anisotropic 
generalisation of the BCS original idea \cite{he3-pwanderson2011}, 
with the same underlying
principle of spontaneous breaking of the symmetry. The search for Majorana
fermions - the particles which are their own antiparticles - 
in exotic superconductors with spin triplet chiral p-wave 
type of order parameter is conducted by several teams of researchers \cite{service2011}. 
The recent connections between the condensed matter physics and string theory
have resulted in the new ideas and ways of approaching
the pairing known as holographic superconductivity. More generally,
the application of holographic methods of string theory 
to condensed matter physics
brings a hope of accessing the strong coupling limit in the models
and theories of the latter branch of physics \cite{hartnoll2009}.

The generalization of the spontaneous symmetry breaking from gauge symmetry to
non-Abelian symmetries of interacting particles has become the most essential
ingredient of the modern theory of elementary particles 
\cite{gautam2011}. The so called
Anderson-Higgs mechanism \cite{anderson1963b,higgs2007} is expected 
to generate masses of particles.
The notion of spontaneous symmetry breaking first encountered 
in the BCS theory of superconductivity has
penetrated to many branches of physics and became one of the paradigms 
of modern physics \cite{schakel2008}.

{\bf Acknowledgmements:} This work has been partially supported 
by the  Ministry of Science and Education under 
the grant No. N202  2631 38. I am grateful to Professor Tadeusz Doma\'nski
for encouraging me to write this paper.

\end{document}